\documentclass[pra,twocolumn,showpacs,preprintnumbers,amsmath,amssymb,superscriptaddress]{revtex4}

\usepackage{bm}
\usepackage{graphicx}
\usepackage{amsbsy}
\usepackage{amsmath}
\usepackage{amsfonts}
\usepackage{amsthm}

\begin{document}

\theoremstyle{plain}
\newtheorem{theorem}{Theorem}
\newtheorem{lemma}[theorem]{Lemma}
\newtheorem{corollary}[theorem]{Corollary}
\newtheorem{conjecture}[theorem]{Conjecture}
\newtheorem{proposition}[theorem]{Proposition}

\theoremstyle{definition}
\newtheorem{definition}{Definition}

\theoremstyle{remark}
\newtheorem*{remark}{Remark}
\newtheorem{example}{Example}

\title{Entanglement of Collaboration}   
\author{Gilad Gour}\email{gour@math.ucalgary.ca}
\affiliation{Institute for Quantum Information Science and 
Department of Mathematics and Statistics,
University of Calgary, 2500 University Drive NW,
Calgary, Alberta, Canada T2N 1N4} 

\date{\today}

\begin{abstract} 
The entanglement of collaboration (EoC) quantifies the maximum amount of entanglement, that can be generated between
two parties, A and B, given collaboration with $N-2$ other parties, when the $N$ parties share a multipartite 
(possibly mixed) state and where the collaboration consists of local operations and classical communication 
(LOCC) by \emph{all} parties. The localizable entanglement (LE) is defined similarly except that A and B
do not participate in the effort to generate bipartite entanglement. We compare between these two operational
definitions and find sufficient conditions for which the EoC is equal to the LE. In particular, we find
that the two are equal whenever they are measured by the concurrence or by one of its generalizations called
the G-concurrence. We also find a simple expression for the LE in terms of the Jamiolkowski isomorphism and
prove that it is convex.
\end{abstract}  


\maketitle

\section*{Introduction}

Entanglement, and in particular, bipartite entanglement has been recognized as a 
valuable resource for important quantum information processing tasks, such as teleportation~\cite{Ben93} 
and superdense coding~\cite{Ben92}. In particular, when a quantum system shared by spatially separated 
parties, entanglement is a resource with which the restriction to local operations and 
classical communications (LOCC) can be overcome. Further restrictions on the amount and/or direction
in which the classical messages are exchanged by the different parties give rise to different
types of entanglement. For example, consider  the distillable 
entanglement~\cite{Ben96}; that is, the amount of Bell states (singlets) that can be
distilled from a bipartite state, $\rho$, in the asymptotic limit of many copies.
If the state $\rho$ is pure then optimal distillation doesn't require any classical 
communication. However, if $\rho$ is mixed then there are at least two types of distillable entanglement
measures corresponding to 1-way and 2-way classical channels.

In this paper, we discuss the difference between 1-way and 2-way classical channels
in the context of entanglement of assistance (EoA)~\cite{Coh98,DiV98,Smo05} and 
localizable entanglement (LE)~\cite{Pop05,Hor05}. 
EoA quantifies the entanglement that can be generated between two parties, 
Alice and Bob, given assistance from a third party, Charlie, when the three share a tripartite state 
and where the assistance consists of Charlie initially performing a measurement on his share and 
communicating the result to Alice and Bob through a one-way classical channel.
After Alice and Bob receive the message from Charlie they end up with more entanglement 
then they had initially.
Thus, we can view this operational definition as a method to \emph{lock} (or more precisely, unlock)
bipartite entanglement
in tripartite states, where Charlie holds the classical key to unlock it. Similarly, the generalization
of EoA to more then three parties, i.e. the LE, can be viewed as locking bipartite entanglement
in multipartite states. In this paper, the term LE refers also to the EoA.  

In~\cite{GR}, a slightly different operational definition has been discussed, dubbed the entanglement
of collaboration (EoC), in which Alice and/or Bob are allowed to perform measurements and 
announce the outcome prior to the measurements performed by Charlie. It has been found that 
with this collaborative tripartite LOCC it is possible to increase the amount of entanglement 
that can be unlocked by Charlie. In what follows, we compare between these two scenarios, and find, 
somewhat surprisingly, that the EoC of a multipartite \emph{mixed} state is \emph{equal} to the LE 
whenever the entanglement between Alice and Bob is measured with the concurrence~\cite{Woo98} or with the 
G-concurrence~\cite{Gou05,Gour05}; this is despite the fact that the
EoC can be strictly greater than the LE when measured with the entropy of entanglement~\cite{Nie00}.
This result is very important since the concurrence have been used widely in the study
of LE, especially, in spin chains~\cite{Pop05,Jin04,Ven05,Sub04,Pac04,Syl03,Ver04}. 

\section*{The localizable entanglement is not an entanglement monotone}

Let us first describe, in a simple way, the example given in~\cite{GR} for which
the EoC is greater than the LE.  
In this example, we consider three parties, Alice, Bob and Charlie sharing an
an $8\times4\times2$ tripartite pure state. The quantum state can be constructed as follows~\cite{RD}: 
consider classical information as a message $y=0,1$ that is encoded in a basis $\{|y\rangle_{C}\}$. 
This classical information can be locked by applying one of two unitaries $\{V_x\}$, where $x=0,1$.
Imagine that Charlie holds the token of the classical message, and Alice holds the key to unlocking 
it. Furthermore, imagine that Alice and Bob share a $4\times 4$ maximally entangled state, and a unitary
on Bob's system is controlled by $x$ and $y$. That is, the state is 
\begin{equation}
|\Psi\rangle=\frac{1}{2}\sum_{x=0}^{1}
\sum_{y=0}^{1}|x\rangle_{a}(I\otimes U_{xy})|\phi^{+}\rangle_{AB}V_{x}|y\rangle_{C}\;,
\label{class}
\end{equation}
where Alice holds systems $a$ and $A$, and 
$|\Phi^{+}\rangle=(|00\rangle+|11\rangle+|22\rangle+|33\rangle)/2$ is the $4\times 4$ maximally
entangled state. The intuition behind this construction is that Charlie holds
the key to getting the entanglement out of Alice and Bob, but it is locked with
information that only Alice can supply. 
Note that if Alice measure $x$ and send the result to Charlie, then Charlie can measure
$y$ and Alice and Bob end up with the maximally entangled state $(I\otimes U_{xy})|\phi^{+}\rangle_{AB}$.
Thus, the EoC is two ebits and it is independent of the choice of the unitaries
$\{V_{x}\}$ and $\{U_{xy}\}$. The LE, on the other hand, does depend on the choice
of $\{V_{x}\}$ and $\{U_{xy}\}$, and can be less than 2 ebits. The pure state example in~\cite{GR}, 
for which the LE (i.e. EoA) is strictly less then 2 ebits, can be written in the same form as in Eq.~(\ref{class})
with, $V_0=I_{2\times 2}$, the identity $2\times 2$ matrix, $V_1=(I_{2\times 2}+\sigma_{y})/\sqrt{2}$,
where $\sigma_{y}$ is the second Pauli matrix, and the four unitaries $U_{xy}$ are diagonal with
$U_{x0}=I_{4\times4}$ for $x=0,1$, and $U_{01}=\text{diag}(i,1,-i,-1)$, $U_{11}=\text{diag}(i,1,i,1)$.
In fact, for this example, it has been shown in~\cite{GR} that Charlie can not create 2 ebits between 
Alice and Bob even with some probability less than one.  

Since the dimension of Alice-Bob system is $8\times 4$, the example above does not rule out
the possibility that the LE is a monotone for lower dimensions or when the entanglement between
Alice and Bob is measured with a monotone that can not distinguish maximally entangled states 
from non-maximally entangled states. Indeed, among other things, we show here that 
for some monotones of this sort, the LE is an entanglement monotone and therefore equal to the 
EoC.

The examples in~\cite{GR} and Eq.~(\ref{class}) also demonstrate the advantage of
2-Way classical channels over 1-Way channels in the process of generating entanglement
between distant parties. To see that, consider a chain of $n$ copies of the  
state~(\ref{class}) shared by $3n$ parties $A_k$, $B_k$ and $C_k$, where $k=1,2,...,n$
and each three parties $A_k$, $B_k$ and $C_k$ share one copy of~(\ref{class}). 
Suppose now that the parties are aligned in a row, such that 
for each $k$, the party $C_k$ is located exactly between $A_k$ and $B_k$, and the parties
$B_k$ and $A_{k+1}$ ($k=1,2,...,n-1$) are close to each other so that they can perform
joint measurements on there spatially separated systems. 
In this scenario, the EoC between the two parties located
at the edges of the chain, i.e. $A_{1}$ and $B_{n}$,
is 2 ebits, since the parties can generate 2 ebits between $A_k$ and $B_k$ and then
use entanglement swapping to generate maximally entangled state between $A_{1}$ and $B_{n}$.
Now, if we add the restriction that no classical information can be transmitted in the direction
from $A_k$ to $C_k$, than it is impossible to generate a maximally entangled state between
$A_k$ and $C_k$ even with some probability. As a result, the maximum average entanglement,  
when measured by the G-concurrence (see Eqs.~(\ref{deta},\ref{CMM})), that can be generated 
between $A_k$ and $B_k$ is bounded above by a positive number $c<1$. 
From the corollary after theorem 1 in~\cite{Gou05}, it follows that the maximum G-concurrence
that can be generated between $A_{1}$ and $B_{n}$ is bounded from above by $c^n$. Therefore, 
in the limit $n\rightarrow\infty$ the maximum G-concurrence
that can be generated between $A_{1}$ and $B_{n}$ approaches zero; this implies that the
entropy of entanglement is bounded by $\log_{2}3$ ebits (because zero G-concurrence implies that
at least one of the Schmidt coefficients is zero). Hence, we can see a significant advantage
of 2-way over 1-way classical channels.  

\section*{Definitions and notations}

The definition of EoC (and LE) corresponds to a family of measures, with each one being parasitic on 
(i.e. defined in terms of) a different bipartite measure of entanglement. The latter is required 
to be an entanglement monotone with respect to LOCC operations on the bipartite system. Following~\cite{GR}, 
we call it the \emph{root entanglement measure} and denote it by $E_{\text{Rt}}.$

\begin{definition}
Given a \emph{mixed} state of $n$ systems, 
the localizable entanglement is defined as the maximum average of the root entanglement measure 
that a distinguished pair of parties ($A$ and $B$) can share after LOCC 
by the \emph{other} $n-2$ parties. 
\end{definition}
Note that since we consider
here mixed multipartite systems, even in the asymptotic limit there are
many possible choices for the root entanglement 
measure~\footnote{In the asymptotic limit of many copies of a \emph{pure} multipartite 
state, $E_{\rm Rt}$ is unique and is taken to be the entropy of entanglement as the 
optimal protocol generates a probability distribution of bipartite \emph{pure} states shared 
by $A$ and $B$.}.
\begin{definition}
Given a \emph{mixed} state of $n$ systems,  
the entanglement of collaboration is defined as the maximum average of the root entanglement measure 
that a distinguished pair of parties ($A$ and $B$) can share after general LOCC 
by \emph{all} the parties (including $A$ and $B$).
\end{definition}

It is clear from the two definitions above that the EoC$\geq$LE with equality
iff the LE is an entanglement monotone. One of the questions we consider in this paper
is for which root entanglement measures EoC$=$LE. We start with some notations.
 
Let $Y\equiv AB,$ be the system onto which entanglement is to be localized,
and $Z\equiv C_1C_2\dots C_{n-2}$ be the system available to the $n-2$ other 
parties that are trying to assist in the distillation. We denote by $\rho^{YZ}$
the mixed multipartite state shared by the $n$ parties.

Recall that a multipartite measure of entanglement $E$ is
an entanglement monotone iff the following two conditions are 
satisfied~\cite{Vid00}:\newline
(1) For any local operation, $\mathcal{E}_{k}$, performed
by one of the parties (in $Y$ or in $Z$)
\begin{equation}
E(\rho ^{YZ})\geq \sum_{k}p_{k}E(\varrho _{k}^{YZ})\;,
\label{cond}
\end{equation}
where $p_{k}\equiv \mathrm{Tr}\left[ \mathcal{E}_{k}(\rho ^{YZ})\right] $
and $\varrho _{k}^{YZ}\equiv \mathcal{E}_{k}(\rho ^{YZ})/p_{k}$.\newline
(2) $E$ is a convex function, that is, $E(\rho )\leq \sum_{k}w_{k}E(\rho
_{k})$ for any ensemble $\{w_{k},\rho _{k}\}$ such that $\rho
=\sum_{k}w_{k}\rho _{k}.$

\section*{A mathematical expression for the localizable entanglement}
 
In order to determine under what conditions the LE, $E_{Loc}$, satisfies the two conditions
above, an explicit expression for $E_{Loc}(\rho^{YZ})$ is needed. This expression is found
with the help of the Jamiolkowski isomorphism. 

\textbf{The Jamiolkowski isomorphism states}~\cite{Jam72}: 
every density operator $\rho
^{YZ}$ is associated with a CP\ map $\mathcal{J}_{\rho }:\mathcal{B}(
\mathcal{H}^{Z})\rightarrow \mathcal{B}(\mathcal{H}^{Y})$ such that $\rho
^{YZ}=\mathcal{J}_{\rho }\otimes \mathcal{I}(\left\vert \psi
^{+}\right\rangle \left\langle \psi ^{+}\right\vert )$ where $\left\vert
\psi ^{+}\right\rangle =\sum_{i}\left\vert i\right\rangle \otimes \left\vert
i\right\rangle \in \mathcal{H}^{Z}\otimes \mathcal{H}^{Z}$ is 
an unnormalized maximally entangled state.
Furthermore, note that for any given operator $\mathcal{A}$,  
$\mathcal{I}\otimes\mathcal{A}\left\vert\psi ^{+}\right\rangle
= \mathcal{A}^{t}\otimes\mathcal{I}\left\vert\psi ^{+}\right\rangle$,
where $\mathcal{A}^{t}$ is the transpose of $\mathcal{A}$.
Given this we deduce the following:

\begin{proposition} 
The localizable entanglement is given by the expression
\begin{equation}
E_{\text{Loc}}(\rho ^{YZ})=\max_{\{Q_{k}^{Z}\}}\sum_{k}p_{k}E_{
\text{Rt}}(\sigma _{k}^{Y})\label{loc}
\end{equation}
where
\begin{eqnarray}
p_{k} &=&\mathrm{Tr}\left[ \rho ^{Z}Q_{k}^{Z}\right]  \\
\sigma _{k}^{Y} &=&\frac{\mathcal{J}_{\rho }(Q_{k}^{Z})}{p_{k}}\label{sig}
\end{eqnarray}
where $\rho^{Z}\equiv\mathrm{Tr}_{Y}\rho ^{YZ}$ and 
$\mathcal{J}_{\rho }:\mathcal{B}(\mathcal{H}^{Z})\rightarrow \mathcal{B
}(\mathcal{H}^{Y})$ is the map associated with $\rho ^{YZ}$ through the
Jamiolkowski isomorphism and where the maximization is over all POVMs 
$\{Q_{k}^{Z}\}$ that can be implemented locally among the parties of $Z$; thus, 
$Q_{k}^{Z}=Q_{k}^{C_1}\otimes Q_{k}^{C_2}
\otimes\cdots\otimes Q_{k}^{C_{n-2}}$.
\end{proposition}

Note that in Eq.~(\ref{sig}) $Q_{k}^{Z}$ should be replaced with $\left(Q_{k}^{Z}\right)^{t}$.
However, since we take the maximization of all possible POVMs we can drop the transposition sign.
With this expression for the LE, we are ready to examine the two conditions (stated above) for 
monotonicity.

\section*{The localizable entanglement is convex}

Intuitively, the LE can not increase if one of the subsystems is discarded or if some of the 
information about the system is lost.
Therefore, we would expect that the LE is a convex function as 
convexity is associated with loss of information~\footnote{It is interesting to note,
however, that it is not always straightforward to equate loss of information with
mixing~\cite{Ple05}.}. 
Indeed, as we show below, the LE 
is convex. Nevertheless, note that for $n=3$, the LE (in this case called EoA) 
is a \emph{concave} function when considered as a \emph{bipartite} measure~\cite{DiV98}.

\begin{proposition} 
The localizable entanglement is a convex function for \emph{any} root
entanglement measure that is an entanglement monotone.
\end{proposition}

\begin{proof} 
We wish to show that if
\begin{equation}
\rho ^{YZ}=\sum_{l}t_{l}\rho _{l}^{YZ},  \label{rhoYZdecomp}
\end{equation}
then
\begin{equation}
E_{\text{Loc}}(\rho ^{YZ})\leq \sum_{l}t_{l}E_{\text{Loc}}(\rho _{l}^{YZ})
\end{equation}
where the LE, $E_{\text{Loc}}$, is given by ({\it cf} Eq.~(\ref{loc}))
\begin{equation}
E_{\text{Loc}}(\rho ^{YZ})=\max_{\{Q_{k}^{Z}\}}\sum_{k}p_{k}E_{\text{Rt}}
\left[ \frac{\mathcal{J}_{\rho }(Q_{k}^{Z})}{p_{k}}\right]   \label{starstar}
\end{equation}
with $E_{\text{Rt}}$ an entanglement monotone on $Y$, 
$p_k\equiv{\rm Tr}\mathcal{J}_{\rho }(Q_{k}^{Z})$
and where $\mathcal{J}_{\rho }:
\mathcal{B}(\mathcal{H}^{Z})\rightarrow \mathcal{B}(\mathcal{H}^{Y})$ 
is the map associated with $\rho ^{YZ}$ through the Jamiolkowski
isomorphism. By Eq.~(\ref{rhoYZdecomp}), $\mathcal{J}_{\rho }=\sum_{l}t_{l}
\mathcal{J}_{\rho _{l}}$ and in particular,
\begin{equation}
\frac{\mathcal{J}_{\rho }(Q_{k}^{Z})}{p_{k}}
=\sum_{l}\left(\frac{t_{l}q_{kl}}{p_k}\right)\frac{\mathcal{J}_{\rho _{l}}(Q_{k}^{Z})}{q_{kl}}\;,
\end{equation}
where $q_{kl}\equiv {\rm Tr}\left[\mathcal{J}_{\rho _{l}}(Q_{k}^{Z})\right]$.
But $E_{\text{Rt}}$ is an entanglement monotone and therefore also a convex
function, so that,
\begin{equation}
E_{\text{Rt}}\left[ \frac{\mathcal{J}_{\rho }(Q_{k}^{Z})}{p_{k}}\right] \leq
\sum_{l}\left(\frac{t_{l}q_{kl}}{p_k}\right)E_{\text{Rt}}
\left[\frac{\mathcal{J}_{\rho _{l}}(Q_{k}^{Z})}{q_{kl}}\right]\;.
\end{equation}
Combining this result with Eq. (\ref{starstar}), we find
\begin{equation}
E_{\text{Loc}}(\rho ^{YZ})\leq \max_{\{Q_{k^{\prime
}}^{Z}\}}\sum_{k}\sum_{l}t_{l}q_{kl}E_{\text{Rt}}\left[ \frac{\mathcal{J}
_{\rho _{l}}(Q_{k}^{Z})}{q_{kl}}\right]\;.
\end{equation}
Note however that the right hand side will only be larger if we maximize
every element of the sum over $l$, thus
\begin{eqnarray}
E_{\text{Loc}}(\rho ^{YZ}) &\leq &\sum_{l}t_{l}\max_{\{Q_{k^{\prime
}}^{Z}\}}\sum_{k}q_{kl}E_{\text{Rt}}\left[ \frac{\mathcal{J}
_{\rho _{l}}(Q_{k}^{Z})}{q_{kl}}\right]\nonumber\\
&=&\sum_{l}t_{l}E_{\text{Loc}}(\rho _{l}^{YZ}),
\end{eqnarray}
where we have made use of the expression~(\ref{loc}) of 
$E_{\text{Loc}}$ for $\rho_{l}^{YZ}$.
\end{proof}

\section*{Sufficient conditions for monotonicity}

The result above shows that the convexity requirement for 
monotonicity is satisfied by the LE for any choice of root 
entanglement measures. Hence, for a given root entanglement 
measure, the LE is an entanglement monotone (and therefore equal
to the EoC) iff the condition given in Eq.~(\ref{cond}) is satisfied.
In the theorem below we find sufficient requirements on a root
entanglement measure that is generating LE=EoC.   

\begin{theorem}\label{T1}
The localizable entanglement $E_{\text{Loc}}$ is an
entanglement monotone whenever the root entanglement measure $E_{\text{Rt}}$
satisfies:\\
(\textbf{i}) Homogeneity of degree 1: 
$E_{\text{Rt}}(c\rho )=cE_{\text{Rt}}(\rho )$, 
where $c$ is a positive real number.\\
(\textbf{ii}) There exist a function $f$ such that

({\it a}) For an arbitrary trace-decreasing completely positive (CP) map,
$\mathcal{E}$,
$$
E_{\text{Rt}}\left[\mathcal{E}\otimes
\mathcal{I}^{B}(\rho ^{AB})\right]\leq f(\mathcal{E})E_{\text{Rt}}(\rho ^{AB})
$$
for all bipartite states $\rho ^{AB}$.

({\it b}) For any set of trace-decreasing CP maps, $\{\mathcal{E}_{j}\}$,
such that the map $\sum_{j} \mathcal{E}_{j}$ is trace-preserving,
$$
\sum_{j}f(\mathcal{E}_{j})\leq 1.
$$
\end{theorem}
\begin{proof}
Given that convexity is already established for the
LE, all we need to show is that
any local operation performed by one of the parties in 
$Y$ or $Z$ can not increase on average the LE. 
Furthermore, note that we need only consider local operations on the 
two subsystems of $Y$ because it is clear, by the definition of the LE, 
that it cannot increase under LOCC on $Z$. Thus, without loss of generality,
it is left to show that any local operation performed on subsystem $A$
cannot increase the LE. The local operation performed on $A$ with outcomes
$\{j\}$ is described by a set of trace-decreasing completely positive maps
$\{\mathcal{E}_{j}\}$. We need to show that
\begin{equation}
E_{\text{Loc}}(\rho ^{YZ})\geq \sum_{j}q_{j}E_{\text{Loc}}(\varrho
_{j}^{YZ}),
\end{equation}
where $q_{j}\equiv \mathrm{Tr}\left[ \mathcal{E}_{j}(\rho ^{YZ})\right] $
and $\varrho _{j}^{YZ}\equiv \mathcal{E}_{j}(\rho ^{YZ})/q_{j}$. Note that in this short
notation, $\mathcal{E}_{j}$, stands for $\mathcal{E}_{j}\otimes\mathcal{I}^{B}\otimes\mathcal{I}^{Z}$,
where $\mathcal{I}^{B}$ and $\mathcal{I}^{Z}$ are the identity maps on $B$ and $Z$.
From the expression given in Eq.(\ref{loc}) for the LE we have
\begin{equation}
\sum_{j}q_{j}E_{\text{Loc}}(\varrho
_{j}^{YZ})=\sum_{j}q_{j}\max_{\{Q_{k^{\prime }}^{Z}\}}
\sum_{k}p_kE_{\text{Rt}}\left[\frac{\mathcal{J}_{\varrho
_{j}}(Q_{k}^{Z})}{p_k}\right]\;. 
\end{equation}
Now, from the Jamiolkowski isomorphism,
\begin{align}
& \mathcal{J}_{\varrho _{j}}\otimes \mathcal{I}^{Z}\left(\left\vert \psi
^{+}\right\rangle \left\langle \psi ^{+}\right\vert\right)
 =\varrho _{j}^{YZ} 
=\frac{1}{q_{j}}\mathcal{E}_{j}(\rho ^{YZ})\nonumber\\
&\;\;\;\;\;\;\;\;\;\;\;\;\;\;\;=\left(\frac{1}{q_{j}}\mathcal{E}_{j}\circ \mathcal{J}_{\rho }\right)\otimes \mathcal{
I}^{Z}\left(\left\vert \psi
^{+}\right\rangle \left\langle \psi ^{+}\right\vert\right)\;,
\end{align}
so that, $\mathcal{J}_{\varrho _{j}}=\frac{1}{q_{j}}\mathcal{E}_{j}\circ
\mathcal{J}_{\rho }$ (because two local super-operators that have the same
image on the maximally entangled state are equivalent). 
Thus, 
\begin{align}
\sum_{j}q_{j}E_{\text{Loc}}(\varrho _{j}^{YZ})
&=\sum_{j}q_j\max_{\{Q_{k^{\prime }}^{Z}\}}
\sum_{k}p_kE_{\text{Rt}}\left[ \frac{\mathcal{E}_{j}\circ \mathcal{J}
_{\rho }(Q_{k}^{Z})}{q_jp_k}\right]\nonumber\\
&=\sum_{j}\max_{\{Q_{k^{\prime }}^{Z}\}}
\sum_{k}p_kE_{\text{Rt}}\left[\frac{\mathcal{E}_{j}\circ \mathcal{J}
_{\rho }(Q_{k}^{Z})}{p_k}\right]\;,
\end{align}
where the last equality follows from the homogeneity assumption~(\textbf{i}).
Now, by assumption~(\textbf{ii}), there exist a function
$f$ such that
\begin{equation}
\sum_{j}q_{j}E_{\text{Loc}}(\varrho
_{j}^{YZ})\leq\sum_{j}\max_{\{Q_{k^{\prime }}^{Z}\}}
\sum_{k}p_{k}f(\mathcal{E}_{j})E_{\text{Rt}}\left[ \frac{\mathcal{J}%
_{\rho }(Q_{k}^{Z})}{p_{k}}\right]\;.
\end{equation}
Hence,
\begin{align}
\sum_{j}q_{j}E_{\text{Loc}}(\varrho _{j}^{YZ})
& \leq\sum_{j}f(\mathcal{E}_{j})\max_{\{Q_{k^{\prime }}^{Z}\}}
\sum_{k}p_{k}E_{\text{Rt}}\left[ \frac{\mathcal{J}
_{\rho }(Q_{k}^{Z})}{p_{k}}\right]\nonumber\\
&=\left[ \sum_{j}f(\mathcal{E}_{j})\right] E_{\text{Loc}}(\rho ^{YZ})\;.
\end{align}
By assumption~(\textbf{ii})
\begin{equation}
\sum_{j}f(\mathcal{E}_{j})\leq 1\;,
\end{equation}
which complete the proof.
\end{proof}

The theorem above shows that the LE equals the EoC
when the root entanglement measure satisfies conditions~(i) and 
(ii). We now show that there exist at least one measure of 
entanglement, the concurrence, that satisfies these conditions. 

\subsection*{The G-concurrence}

Originally, the concurrence has been defined for a pair of 
qubits~\cite{Woo98}, though generalization to higher dimensions 
are possible~\cite{Run01}, but not unique~\cite{Gou05}. 
Here we focus on one of the generalizations which we call the 
G-concurrence~\cite{Gou05,Gour05}.
  
For a pure bipartite state, $|\psi\rangle$, the G-concurrence is defined as 
the {\em geometric mean} of the (non-negative) Schmidt numbers
\begin{equation} 
G(|\psi\rangle)\equiv 
d(\lambda_{0}\lambda_{1}\cdots\lambda_{d-1})^{\frac{1} {d}}
=d\left[{\rm Det}\left(A^{\dag}A\right)\right]^{\frac{1} {d}}\;,
\label{deta}
\end{equation}
where the matrix elements of $A$ are $a_{ij}$ 
($|\psi\rangle=\sum_{ij}a_{ij}|i\rangle|j\rangle$).
For a mixed $d\times d$-dimensional bipartite state, $\rho$, 
the G-concurrence is defined 
in terms of the convex roof extension:
\begin{equation}
G(\rho)={\rm min}\sum_{i}p_{i}G(|\psi _{i}\rangle)\;\;\left(\rho
=\sum_{i}p_{i}|\psi _{i}\rangle\langle\psi _{i}|\right)\;,
\label{CMM}
\end{equation}
where the minimum is taken over all decompositions of $\rho$. 
In~\cite{Gou05} it has been shown that the G-concurrence as defined in 
Eqs.~(\ref{deta},\ref{CMM}) is a bipartite entanglement monotone.
It has been shown that it can be interpreted operationally as a kind
of entanglement capacity and that it is a computationally manageable measure of 
entanglement~\cite{Gour05}. Very recently, the G-concurrence has been proved 
fruitful in calculating the average entanglement of random bipartite 
pure states~\cite{Cap06} and also played a crucial role in a demonstration
of an asymmetry of quantum correlations~\cite{Hor06}. 
  
\begin{corollary}\label{cor}
The LE is equal EoC whenever the root entanglement measure is taken
to be the G-concurrence.
\end{corollary}

The corollary above is surprising taking into account the example
in~\cite{GR} (\emph{cf} Eq.~(\ref{class})). Note however that the 
corollary does not contradict the results in~\cite{GR}
since the G-concurrence vanish for $m\times n$ states
with $m\neq n$. Furthermore, note that if the parties A and B each hold 
a qubit then the corollary states that the LE, which has 
been studied in~\cite{Pop05} in terms of the concurrence and
in~\cite{Jin04,Ven05,Sub04,Pac04,Syl03,Ver04} for spin chains, is 
indeed an entanglement monotone.

\begin{proof} (of corollary~\ref{cor})

The corollary above follows from two properties of the G-concurrence~\cite{Gou05}: 
\begin{align}
G(c|\psi\rangle) & = |c|^{2}G_{d}(|\psi\rangle)\label{m1}\\
G\left(\hat{A}\otimes\hat{B}|\psi\rangle\right) & =
\left|{\rm Det}\left(\hat{A}\right)\right|^{2/d}\left|{\rm Det}\left(\hat{B}\right)
\right|^{2/d}G(|\psi\rangle)\;.
\label{pro}
\end{align}   
Hence, condition~(i) in theorem~\ref{T1} is satisfied, as the G-concurrence is homogeneous.
We now show that also condition~(ii) is satisfied.

Given a bipartite density matrix $\rho^{AB}=\sum_i p_i |\psi_i\rangle\langle\psi _i|$
we take $\{p_i,\;|\psi_i\rangle\}$ to be an optimal decomposition such that
$$
G(\rho^{AB})=\sum_i p_i G(|\psi_i\rangle)\;.
$$
Let $\{\mathcal{E}_j\}$ be a set of trace-decreasing CP maps,
such that the map $\sum_{j} \mathcal{E}_{j}$ is trace-preserving.
Each map can then be written in the Kraus form:
$$
\mathcal{E}_{j}(\rho^{AB})=\sum_{k}M_{jk}^\text{A}\otimes\mathcal{I}^\text{B}
\rho^{AB} M_{jk}^{\text{A}\dag}\otimes\mathcal{I}^\text{B}\;,
$$
where $\sum_{jk}M_{jk}^{\dag}M_{jk}=\mathcal{I}^\text{A}$.
Now, we denote the normalized state 
$|\phi_{jki}\rangle\equiv N_{ikj}^{-1/2}  M_{jk}\otimes\mathcal{I}|\psi_i\rangle$, where
the normalization factor, $N_{ikj}$, is taken such that $\langle\phi_{jki}|\phi_{jki}\rangle=1$.
Thus, 
$$
\mathcal{E}_{j}(\rho^{AB})=\sum_{i,k}p_iN_{ikj}|\phi_{jki}\rangle\langle\phi_{jki}|\;,
$$
and since $G$ is defined in terms of the convex roof extension we have
\begin{align}
& G\left(\mathcal{E}_{j}(\rho^{AB})\right)\leq\sum_{i,k}p_iN_{ikj}G(|\phi_{jki}\rangle)\nonumber\\
& =\sum_{i,k}p_iG(M_{jk}\otimes\mathcal{I}||\psi_i\rangle)
=\sum_{k}|\text{Det}M_{jk}|^{2/d}\;G(\rho^\text{AB})\;,
\nonumber
\end{align}
where we have used the two properties given in Eq.~(\ref{pro})
and the optimality of the decomposition $\{p_i,\;|\psi_i\rangle\}$. 
Hence, defining
$f(\mathcal{E}_{j})\equiv \sum_{k}|\text{Det}M_{jk}|^{2/d}$ we get 
$G\left(\mathcal{E}_{j}(\rho^{AB})\right)\leq f(\mathcal{E}_{j})G(\rho^\text{AB})$
and
$$
\sum_{j}f(\mathcal{E}_{j})=\sum_{j,k}|\text{Det}M_{jk}|^{2/d}
\leq \frac{1}{d}\sum_{j,k}\text{Tr}M_{jk}^{\dag}M_{jk}=1\;,
$$
where we have used the geometric-arithmetic inequality. This conclude the proof.
\end{proof}

\section*{Summary and conclusions}

In summary, following~\cite{GR}, we have introduced the EoC
and compared it with the LE. With the help of the Jamiolkowski 
isomorphism, we were able to find a simple expression for the LE
of multipartite mixed states and to prove that it is a convex 
function. We have also found a set of sufficient conditions for 
which the EoC equals the LE. We have shown that these conditions 
are met when the LE and the EoC are measured with the concurrence 
or with one of its generalizations to higher dimensions called 
the G-concurrence.

The example given in~\cite{GR} (\emph{cf} Eq.~(\ref{class})) shows that the 
LE fail to be an entanglement monotone (the usual kind, with respect to unrestricted 
LOCC) when it is defined in terms of a root entanglement measure that can
distinguish maximally entangled states from non-maximally entangled
states. This left open the possibility that the LE can
be an entanglement monotone for choices of the root measure that are
not of this sort. Here we have shown that this possibility is
indeed realized in the case where the root measure is one of 
the generalizations of the concurrence. 
Determining in which Hilbert spaces and for which root
entanglement measures the LE is a monotone will help to identify
those distributed QIP tasks for which having a collaboration among
all parties provides no advantage over merely having the assistance
of $n-2$ parties.

\emph{Acknowledgments:---}
The author would like to thank Rob Spekkens for many fruitful
discussions.

\end{document}